# Simulating the nanomechanical response of cyclooctatetraene molecules on a graphene device

Sehoon Oh[1,2], Michael F. Crommie[1,2,3] and Marvin L. Cohen[1,2]*

[1]*Department of Physics, University of California at Berkeley, Berkeley, CA 94720, USA*

[2]*Materials Sciences Division, Lawrence Berkeley National Laboratory, Berkeley, CA 94720, USA*

[3]*Kavli Energy Nano Sciences Institute at the University of California Berkeley and the Lawrence Berkeley National Laboratory, Berkeley, CA 94720, USA*

* email : mlcohen@berkeley.edu

**ABSTRACT** : We investigate the atomic and electronic structures of cyclooctatetraene (COT) molecules on graphene and analyze their dependence on external gate voltage using first-principles calculations. The external gate voltage is simulated by adding or removing electrons using density functional theory (DFT) calculations. This allows us to investigate how changes in carrier density modify the molecular shape, orientation, adsorption site, diffusion barrier, and diffusion path. For increased hole doping COT molecules gradually change their shape to a more flattened conformation and the distance between the molecules and graphene increases while the diffusion barrier drastically decreases. For increased electron doping an abrupt transition to a planar conformation at a carrier density of $-8\times10^{13}$ e/cm$^2$ is observed. These calculations imply that the shape and mobility of adsorbed COT molecules can be controlled by externally gating graphene devices.

**Table of contents graphic**

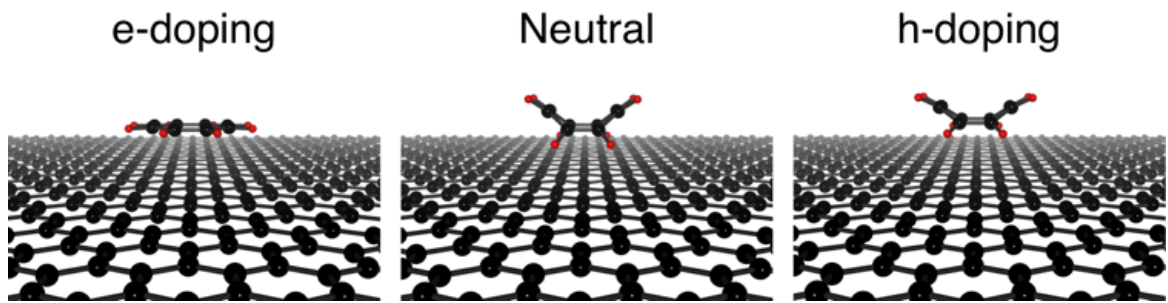

COT molecules are the first known $4n$ π-electron hydrocarbon system [1,2]. Unlike benzene, COT adopts a nonplanar tub-shaped conformation of $D_{2d}$ symmetry with alternating single and double bonds [3-5] so that the interaction between the double bonds is minimized [5,6]. It has been observed that the molecule undergoes thermal bond shift and ring inversion processes [6-8]. The ring inversion involves a planar transition state of $D_{4h}$ symmetry having alternate single and double bonds while the bond shift proceeds via a planar transition state of $D_{8h}$ symmetry in which all C-C bonds have equal length [9,10]. In contrast, COT's dianion, cyclooctatetraenide (COT$^{2-}$), adopts a planar conformation of $D_{8h}$ symmetry according to NMR experiments. This pronounced coupling between the COT charge state and its mechanical conformation creates new opportunities for exploiting COT as a single-molecule electromechanical transducer element. The COT charge state, for example, might be switched by attaching it to a gate-tunable graphene field-effect transistor (FET), thus triggering a mechanical response. Graphene FETs are ideal for this application since they allow both electron and hole carrier densities up to $|\rho| \sim 10^{13}$ e/cm$^2$ when gate voltages are applied to graphene on an insulating layer (e.g., SiO$_2$ or BN) [11,12]. Even higher carrier densities up to $|\rho| \sim 10^{14}$ e/cm$^2$ are possible with electrochemical gating [13-16]. This allows the charge of molecules adsorbed onto graphene to be precisely controlled by adjusting applied external gate voltages, as has been seen previously for different molecular systems that have no nanomechanical functionality [17-22].

In order to explore the potential nanomechanical functionality of COT, we have theoretically investigated the gate voltage dependence of the atomic and electronic structures of COT molecules adsorbed onto graphene through the use of *ab initio* density functional theory (DFT). External gating of COT is simulated by changing the graphene carrier density in our DFT calculations. As shown below, COT molecules gradually flatten as the system is positively charged, while they abruptly change shape to a planar conformation at a carrier density $\rho = -8 \times 10^{13}$ e/cm$^2$ when negatively charged, where $-e$ is the electron charge. We show that the diffusivity of a COT molecule, as well as its conformation, can be controlled by the application of achievable gate voltages.

This paper is organized as follows: we first review our calculational methods and then report on the atomic and electronic structures of freestanding COT molecules. We next discuss the properties of COT molecules adsorbed onto graphene, and the carrier density dependence of COT properties is analyzed.

Our calculations were performed within the framework of pseudopotential DFT (density functional theory). We used the generalized gradient approximation [23], norm-conserving pseudopotentials [24], and localized pseudoatomic orbitals for wavefunctions as implemented in the SIESTA code [25]. The real-space mesh cut-off of 1000 Ry is used for all of our calculations. Dipole corrections are included to reduce the interactions between the supercells generated by the periodic boundary condition [26]. The atomic positions are fully optimized until residual forces are less than $10^{-3}$ eV/Å. During the atomic structure optimization, a DFT-D2 correction is applied to account for the van der Waals interaction [27].

First, we investigate the atomic and electronic structures of a freestanding COT molecule using a DFT calculation for a cubic cell of length 20 Å with only the Γ point of the Brillouin zone sampled for the molecule. The neutral COT molecule and its ionized states are studied. For each ionized states, three conformations, a nonplanar tub-shaped conformation of $D_{2d}$ symmetry with alternating single and double bonds [Fig. 1(A)], a planar conformation of $D_{4h}$ symmetry with alternating single and double bonds [Fig. 1(B)], and a planar cyclic conjugated conformation of $D_{8h}$ symmetry in which all of C-C bonds are equal in length [Fig. 1(C)], are investigated. We optimize the atomic positions of the three conformations having $D_{2d}$, $D_{4h}$, and $D_{8h}$ symmetries by minimizing the total energy with the symmetry constraints. The obtained structural parameters of the neutral molecule, dication ($COT^{2+}$), monovalent cation ($COT^{+}$), monovalent anion ($COT^{-}$), and dianion ($COT^{2-}$) are presented in Tab. 1. As presented in Tab. 1, COT molecules adopt different symmetry conformations depending on their ionized states. The neutral molecule, the monovalent cation and the dication adopt nonplanar tub-shaped conformations of $D_{2d}$ symmetry, while monovalent anion and the dianion adopt planar conformations of $D_{4h}$ and $D_{8h}$ symmetry, respectively.

The neutral molecule energetically favors a tub-shaped conformation of $D_{2d}$ symmetry with the bent angle of the COT ring α=41.8°, in agreement with the previous studies. The energy difference between the highest occupied molecular orbital (HOMO) and lowest unoccupied molecular orbital (LUMO) states $E_g$ is 2.69 eV. The energies of the transition states with $D_{4h}$ and $D_{8h}$ conformations are 441.6 and 870.1 meV higher than the ground state with $D_{2d}$ conformation, respectively. The monovalent cation adopts a tub-shaped conformation with α=30.0° and $E_g$=1.19 eV. The transition states of monovalent cation of $D_{4h}$ and $D_{8h}$ conformations respectively have 91.3 and 212.9 meV higher energy than the ground state of $D_{2d}$ conformation. For dication, the ground state of $D_{2d}$ conformation with α=10.9° has 1.2 meV lower energy than the transition state of the planar conformation, which means that the ring inversion and the bond shift are prevalent at room temperature. In the case of the monovalent anion, the $D_{4h}$ conformation is the ground state having 102.3 meV higher energy than the $D_{8h}$ conformation due to Jahn Teller distortion. The dianion has the ground state of $D_{8h}$ conformation, which has a pair of filled degenerate non-bonding orbitals, because of the unusual stability of cyclic delocalization of the $4n+2$ π-electron system [28], which is known as aromaticity.

We investigate the charge dependence of the properties further by varying the charge of the molecule $q_{COT}$ as shown in Fig. 2. Although a fractional $q_{COT}$ for a freestanding molecule is not a realistic situation, it helps to analyze the charge dependence of the properties and would be realistic when the COT molecule is adsorbed onto graphene with an external gate voltage. As positively charged, a COT molecule is gradually flattened toward the $D_{8h}$ conformation [Fig. 2(A)], and the energy barriers for the ring inversion $E_{RI}=E_{D2d}-E_{D4h}$ and the bond shift $E_{BS}=E_{D4h}-E_{D8h}$ decrease [Fig. 2(B)], where $E_{D2d}$, $E_{D4h}$, and $E_{D8h}$ are the total energies of $D_{2d}$, $D_{4h}$, and $D_{8h}$ conformations, respectively. For a negatively charged case, α decreases rapidly as the molecule is charged, and the molecule is completely flattened with a $D_{4h}$ conformation at $q_{COT}=-0.7$ e, and becomes a $D_{8h}$ conformation at $q_{COT}=-2.0$ e. Figure 2(C) shows the charge dependence of $E_g$. As the molecule is positively charged, $E_g$ gradually decreases as the conformation goes toward $D_{8h}$ conformation, whereas it decreases rapidly as going toward $D_{4h}$ conformation and gradually as going toward $D_{8h}$ conformation with negative charge. Figure 2(D) shows the relationship between $E_g$ and α. As α decreases, $E_g$ decreases considerably, consistent with previous study [9]. When the molecule is completely planarized, $E_g$ remains finite for the neutral molecule and the monovalent ions because they have $D_{4h}$ conformation, while becomes zero for the divalent ions because of $D_{8h}$ symmetry.

We next investigate the atomic and electronic structures of a COT molecule adsorbed onto graphene. DFT calculations are performed using a periodic hexagonal 5×5 supercell of the graphene unitcell with one adsorbed COT molecule on it. This supercell contains 66 atoms, 8 carbon atoms and 8 hydrogen atoms of the molecule, and 50 carbon atoms of graphene. The vacuum region in the supercell is ~50 Å thick along the direction perpendicular to the graphene layer to avoid fictitious interactions generated by the periodic boundary conditions. The Brillouin zone is sampled with 32×32 Monkhorst-Pack k-point mesh [29]. We construct initial candidate structures using the atomic positions of a freestanding COT molecule with three different symmetries and pristine graphene with various adsorption sites and orientations of the molecule. As for the adsorption site of a COT molecule, we consider various site including three high-symmetry sites, in which the center of mass of the molecule is vertically located right on the center of a hexagon of graphene (hollow), on the midpoint of a π-bond of graphene (bridge), on top of a carbon atom of graphene (top), as schematically shown in Fig. 3(B). Various orientations of the molecule are also considered [Fig. S2]. From the constructed initial candidate structures, the atomic positions of all atoms in the supercell are optimized by minimizing the total energy to find the ground state and transition state geometries.

The properties of the molecule on graphene under external gate voltage are also investigated by adding or removing electrons in the DFT calculation. We calculate the properties with 12 charged configurations, $\pm 8.823 \times 10^{13}$, $\pm 7.352 \times 10^{13}$, $\pm 5.882 \times 10^{13}$, $\pm 4.441 \times 10^{13}$, $\pm 2.941 \times 10^{13}$, $\pm 1.470 \times 10^{13}$ e/cm$^{-2}$, corresponding to 1.2, 1.0, 0.8, 0.6, 0.4, 0.2 electrons removed and added to the cell, respectively, + for electron removed and − for added. We calculate the charge transfer to the molecule using the Mülliken population analysis [30]. We obtain the binding energy of a COT molecule $E_{bind}$ as a function of the carrier density $\rho$, which is obtained by $E_{bind}(\rho) = E_{grap}(\rho) + E_{COT}$

− $E_{\text{COT+grap}}(\rho)$, where terms on the right-hand side are the total energies of graphene with carrier density $\rho$, a freestanding neutral COT molecule of $D_{2d}$ conformation, and COT adsorbed onto graphene with carrier density $\rho$, respectively. The diffusion barrier height $E_d$ is assumed to be the difference of the total energies of the ground state (GS) and the lowest energy transition state (LTS) geometries, $E_d(\rho) = E_{\text{LTS}}(\rho) - E_{\text{GS}}(\rho)$, where the $E_{\text{LTS}}(\rho)$ and $E_{\text{GS}}(\rho)$ are the total energies of LTS and GS with carrier density $\rho$, respectively. The overall gate voltage dependence of the diffusion barrier can be inferred from the obtained $E_d(\rho)$, although it would require more sophisticate methods such as the nudged elastic band method [31] to determine a more accurate diffusion barrier. We assume that the most likely diffusion pathway between the adsorption sites of GS is via that of LTS.

For a neutral system (without external gating), the GS geometry is shown in Fig. 4(C). The COT molecule is adsorbed at a hollow site, and the pair of double bonds on the upper plane of the molecule is parallel to a π-bond of graphene. The adsorbed molecule exhibits a nonplanar tub-shaped conformation similar to the neutral freestanding molecule except that the adsorbed molecule is slightly flatter than the freestanding molecule. The adsorbed molecule has 1.63° smaller α than the freestanding molecule. The distance between the molecule and graphene, $d$, is 2.519 Å. The adsorption of the molecules hardly changes the atomic structure of graphene and does not alter the energy levels of the molecular orbital states significantly as shown in Fig. 4(H). No significant covalent bonding between the molecule and graphene is found. The energy level of the Dirac points of graphene, $E_{\text{Dirac}}$, lies between the energy levels of HOMO and LUMO states, $E_{\text{HOMO}}$ and $E_{\text{LUMO}}$, which are 0.70 eV lower and 1.85 eV higher than $E_{\text{Dirac}}$, respectively. As a result, no significant charge transfer occurs between the molecule and graphene, and the Fermi energy $E_F$ is equal to $E_{Dirac}$ as shown in Fig. 4(H). The calculated binding energy is 704 meV and the diffusion barrier is 28.4 meV. The diffusion path is via a location between bridge and top sites.

Next, we study the carrier density dependence of the atomic and electronic properties of COT molecules adsorbed onto graphene. Figure 4 shows the overall carrier density dependence of the atomic and electronic band structures of COT molecules adsorbed onto graphene. The molecule is gradually flattened as the hole carrier density increases, while changes abruptly its shape to a nearly planar conformation with the electron carrier density. The distance $d$ increase with the hole carrier density, while decrease with the electron carrier density. The obtained structural parameters, binding energy, energy barriers of the COT molecule, and charge transfer to the molecule $\Delta q_{COT}$ with various $\rho$ are listed on Tab. 2.

For hole-doped case, the Fermi energy $E_F$ decreases as $\rho$ increases from $\rho = 0$, where $E_F = E_{Dirac}$, down to $E_{HOMO}$, until $\rho$ reaches $\rho_{h1} = +1.5 \times 10^{13}$ e/cm$^2$, where $E_F = E_{HOMO}$. As the hole carrier density increases further beyond $\rho_{h1}$, the molecule is positively charged as the HOMO states becomes partially occupied. Figure 5(A) shows that the charge transferred to the molecule $\Delta q_{COT}$ increases linearly with the hole carrier density when $\rho > \rho_{h1}$. Consequently, the molecule is flattened [Fig. 5(B)], floats upward [Fig. 5(C)], and the diffusion barrier $E_d$ decreases [Fig. 5(D)] until $\rho$ reaches at $\rho_{h2} = +8.5 \times 10^{13}$ e/cm$^2$, where $E_d$ becomes zero. For $0 < \rho < \rho_{h2}$, the molecule adsorbed at hollow site is the GS, and the adsorption site in LTS is between bridge and top sites. For $\rho > \rho_{h2}$, the molecule is adsorbed between bridge and top sites is the GS, and at a hollow site in LTS.

In the case of the electron-doping, $E_F$ increases as $\rho$ decreases from $\rho = 0$, where $E_F = E_{Dirac}$, up to $E_{LUMO}$, until $\rho$ reaches $\rho_{e1} = -6 \times 10^{13}$ e/cm$^2$, where $E_F = E_{LUMO}$. For $\rho_{e1} < \rho < 0$, the molecule is hardly charged as shown in Figure 5(a) because $E_F$ lies between $E_{HOMO}$ and $E_{LUMO}$, consequently the conformation hardly changes as shown in Fig. 5(B). The molecule slightly comes closer to graphene [Fig. 5(C)], and the diffusion barrier increases [Fig. 5(D)] until $\rho$ reaches at $\rho_{e1}$. The adsorption site is a hollow site in GS, while a location between bridge and top sites in LTS for all the electron carrier density investigated. As $\rho$ decreases further below $\rho_{e1}$, the molecule is negatively charged as the LUMO states becomes occupied. For $\rho < \rho_{e2} = -8 \times 10^{13}$ e/cm$^2$, the molecule adopts a nearly planar $D_{4h}$ conformation as shown in Fig. 4(A). As the molecule is planarized, $d$ increases and $E_d$ is reduced drastically.

We show that it is possible to control the conformation and the diffusivity of COT molecules adsorbed on graphene by changing the carrier density of the system. But the carrier density required to manipulate the conformation and diffusivity of COT is rather high because $E_{HOMO}$ and $E_{LUMO}$ are respectively 0.70 and 1.85 eV apart from $E_{Dirac}$. We suggest substitutions of hydrogen atoms to fluorine, bromine atoms or other electrophiles so that $E_{HOMO}$ and/or $E_{LUMO}$ are closer to $E_{Dirac}$, which makes it easier to manipulate the shape and diffusivity of the COT molecule with lower carrier density. Investigating the steric effects of the substituents would also improve the controllability of mechanical properties, but it is beyond the scope of this paper.

In conclusion, we have investigated the atomic and electronic properties of COT molecule and its dependence of gate voltage. Without extra carrier, the molecule adsorbed at a hollow site of graphene in a tub-shaped conformation without significant deformation and the molecule remains almost neutral because the Dirac points lie between the energy levels of HOMO and LUMO states. When an external gate voltage is applied, the molecule is flattened and the diffusion barrier decreases gradually with the hole carrier, while the molecule exhibits abrupt change of its shape to a nearly planar conformation and the diffusion barrier also rapidly decreases with the electron carrier. With achievable carrier density, it is possible to manipulate the conformation as well as the diffusion barrier of the molecule on graphene. We thus envision transformable and mobile molecular machine on graphene, which can be controlled by adjusting external gate voltage. It opens new avenues for exciting electromechanical applications. We suggest substitutions of hydrogen atoms to fluorine, bromine atoms or other electrophiles so that the energy levels of HOMO and/or LUMO are closer to Dirac point of graphene, which makes it easier to control the mechanical properties of the molecule with lower carrier density.

## ACKNOWLEDGEMENTS

Research supported by the U.S. Department of Energy (DOE), Office of Science, Basic Energy Sciences (BES) under contract no. DE-AC02-05CH11231 within the Nanomachine program (COT DFT simulations), and by the National Science Foundation under Grant No. DMR-1508412 (development of theory formalism).

## RERERENCES

(1) Willstätter, R.; Waser, E. Über Cyclo-Octatetraen. *Chem. Ber.* **1911**, *44*, 3423–3445.
(2) Willstätter, R.; Heidelberger, M. Zur Kenntnis Des Cyclo-Octatetraens. *Chem. Ber.* **1913**, *46*, 517–527.
(3) Karle, I. L. An Electron Diffraction Investigation of Cyclooctatetraene and Benzene. *The Journal of Chemical Physics* **1952**, *20*, 65–70.
(4) Bastiansen, O.; Hedberg, L.; Hedberg, K. Reinvestigation of the Molecular Structure of 1,3,5,7-Cyclooctatetraene by Electron Diffraction. *The Journal of Chemical Physics* **1957**, *27*, 1311–1317.
(5) Traetteberg, M. The Molecular Structure of 1,3,5,7-Cyclo-Octatetraene. *Acta Chem. Scand.* **1966**, *20*, 1724–1726.
(6) Wu, J. I.; Fernández, I.; Mo, Y.; Schleyer, P. V. R. Why Cyclooctatetraene Is Highly Stabilized: the Importance of "Two-Way" (Double) Hyperconjugation. *J. Chem. Theory Comput.* **2012**, *8*, 1280–1287.
(7) Naor, R.; Luz, Z. Bond Shift Kinetics in Cyclo-Octatetraene by Dynamic NMR in Liquid Crystalline Solvents. *The Journal of Chemical Physics* **1982**, *76*, 5662–5664.
(8) Andrés, J. L.; Castaño, O.; Morreale, A.; Palmeiro, R.; Gomperts, R. Potential Energy Surface of Cyclooctatetraene. *The Journal of Chemical Physics* **1998**, *108*, 203–207.

(9) Nishinaga, T.; Ohmae, T.; Iyoda, M. Recent Studies on the Aromaticity and Antiaromaticity of Planar Cyclooctatetraene. *Symmetry* **2010**, *2*, 76–97.
(10) Gellini, C.; Salvi, P. R. Structures of Annulenes and Model Annulene Systems in the Ground and Lowest Excited States. *Symmetry* **2010**, *2*, 1846–1924.
(11) Novoselov, K. S.; Geim, A. K.; Morozov, S. V.; Jiang, D.; Zhang, Y.; Dubonos, S. V.; Grigorieva, I. V.; Firsov, A. A. Electric Field Effect in Atomically Thin Carbon Films. *Science* **2004**, *306*, 666–669.
(12) Novoselov, K. S.; Geim, A. K.; Morozov, S. V.; Jiang, D.; Katsnelson, M. I.; Grigorieva, I. V.; Dubonos, S. V.; Firsov, A. A. Two-Dimensional Gas of Massless Dirac Fermions in Graphene. *Nature* **2005**, *438*, 197–200.
(13) Das, A.; Pisana, S.; Chakraborty, B.; Piscanec, S.; Saha, S. K.; Waghmare, U. V.; Novoselov, K. S.; Krishnamurthy, H. R.; Geim, A. K.; Ferrari, A. C.; *et al.* Monitoring Dopants by Raman Scattering in an Electrochemically Top-Gated Graphene Transistor. *Nature Nanotechnology* **2008**, *3*, 210–215.
(14) Efetov, D. K.; Kim, P. Controlling Electron-Phonon Interactions in Graphene at Ultrahigh Carrier Densities. *Phys. Rev. Lett.* **2010**, *105*, 256805–4.
(15) Ye, J. T.; Inoue, S.; Kobayashi, K.; Kasahara, Y.; Yuan, H. T.; Shimotani, H.; Iwasa, Y. Liquid-Gated Interface Superconductivity on an Atomically Flat Film. *Nat Mater* **2009**, *9*, 125–128.
(16) Bhattacharya, A.; Eblen-Zayas, M.; Staley, N. E.; Huber, W. H.; Goldman, A. M. Micromachined SrTiO3 Single Crystals as Dielectrics for Electrostatic Doping of Thin Films. *Appl. Phys. Lett.* **2004**, *85*, 997–999.
(17) Chen, J. H.; Jang, C.; Adam, S.; Fuhrer, M. S.; Williams, E. D.; Ishigami, M. Charged-Impurity Scattering in Graphene. *Nature Physics* **2008**, *4*, 377–381.
(18) Suarez, A. M.; Radovic, L. R.; Bar-Ziv, E.; Sofo, J. O. Gate-Voltage Control of Oxygen Diffusion on Graphene. *Phys. Rev. Lett.* **2011**, *106*, 141–144.
(19) Wang, X.; Xu, J.-B.; Xie, W.; Du, J. Quantitative Analysis of Graphene Doping by Organic Molecular Charge Transfer. *J. Phys. Chem. C* **2011**, *115*, 7596–7602.
(20) Torsten Hahn, S. L. J. K. A Gate Controlled Molecular Switch Based on Picene–F4TCNQ Charge-Transfer Material. *Nanoscale* **2014**, *6*, 14508–14513.
(21) Tsai, H.-Z.; Omrani, A. A.; Coh, S.; Oh, H.; Wickenburg, S.; Son, Y.-W.; Wong, D.; Riss, A.; Jung, H. S.; Nguyen, G. D.; *et al.* Molecular Self-Assembly in a Poorly Screened Environment: F 4TCNQ on Graphene/BN. *ACS Nano* **2015**, *9*, 12168–12173.
(22) Riss, A.; Wickenburg, S.; Tan, L. Z.; Tsai, H.-Z.; Kim, Y.; Lu, J.; Bradley, A. J.; Ugeda, M. M.; Meaker, K. L.; Watanabe, K.; *et al.* Imaging and Tuning Molecular Levels at the Surface of a Gated Graphene Device. *ACS Nano* **2014**, *8*, 5395–5401.
(23) Perdew, J. P.; Burke, K.; Ernzerhof, M. Generalized Gradient Approximation Made Simple. *Phys. Rev. Lett.* **1996**, *77*, 3865–3868.
(24) Troullier, N.; Martins, J. L. Efficient Pseudopotentials for Plane-Wave Calculations. *Phys. Rev. B* **1991**, *43*, 1993–2006.
(25) Soler, J. M.; Artacho, E.; Gale, J. D.; Garcia, A.; Junquera, J.; Ordejon, P.; Sanchez-Portal, D. The SIESTA Method for Ab Initio Order-N Materials Simulation. *Journal of Physics: Condensed Matter* **2002**, *14*, 2745–2779.
(26) Cohen, M. L.; Schlüter, M.; Chelikowsky, J. R.; Louie, S. G. Self-Consistent Pseudopotential Method for Localized Configurations: Molecules. *Phys. Rev. B* **1975**, *12*, 5575–5579.

(27) Grimme, S. Semiempirical GGA-Type Density Functional Constructed with a Long-Range Dispersion Correction. *Journal of Computational Chemistry* **2006**, *27*, 1787–1799.

(28) Hückel, E. Quantum-Theoretical Contributions to the Benzene Problem. *Z. Phys* **1931**, *70*, 204–286.

(29) Monkhorst, H. J.; Pack, J. D. Special Points for Brillouin-Zone Integrations. *Phys. Rev. B* **1976**, *B13*, 5188–5192.

(30) Mulliken, R. S. Electronic Population Analysis on LCAO–MO Molecular Wave Functions. I. *The Journal of Chemical Physics* **1955**, *23*, 1833–1840.

(31) Henkelman, G.; Uberuaga, B. P.; Jónsson, H. A Climbing Image Nudged Elastic Band Method for Finding Saddle Points and Minimum Energy Paths. *The Journal of Chemical Physics* **2000**, *113*, 9901–9904.

**Tab. 1. Structural properties of a freestanding neutral COT molecule and its ionized states.** Symmetry of the molecule, the bond lengths between carbon atoms, $d_{CC1}$ and $d_{CC2}$, between carbon and hydrogen atoms, $d_{CH}$, the distance between the planes of the upper and bottom four carbon atoms of the molecule, $d_{z1}$, the upper and bottom four hydrogen atoms of the molecule, $d_{z2}$, and the bent angle, $\alpha$, as denoted in Fig. 1(A), of the neutral molecule, the monovalent ions, and the divalent ions are presented.

| | Symmetry | $d_{CC1}$(Å) | $d_{CC2}$(Å) | $d_{CH}$(Å) | $d_{z1}$(Å) | $d_{z2}$(Å) | $\alpha(°)$ |
|---|---|---|---|---|---|---|---|
| $COT^{2+}$ | $D_{2d}$ | 1.438 | 1.438 | 1.121 | 0.194 | 0.412 | 10.9 |
| $COT^{+}$ | $D_{2d}$ | 1.400 | 1.463 | 1.119 | 0.553 | 1.259 | 30.0 |
| COT | $D_{2d}$ | 1.373 | 1.491 | 1.120 | 0.797 | 1.972 | 41.8 |
| $COT^{-}$ | $D_{4h}$ | 1.408 | 1.465 | 1.121 | 0.000 | 0.000 | 0.000 |
| $COT^{2-}$ | $D_{8h}$ | 1.444 | 1.444 | 1.128 | 0.000 | 0.000 | 0.000 |

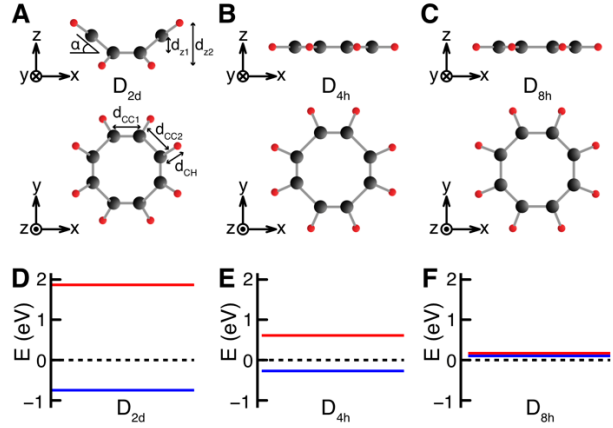

**Fig. 1. Atomic and electronic structures of a freestanding COT molecule.** (A-C) Atomic structures of COT molecule with (A) $D_{2d}$, (B) $D_{4h}$, and (C) $D_{8h}$ symmetry are schematically shown. Side and top views are presented in the top and bottom rows, respectively. Black and red spheres represent carbon and hydrogen atoms, respectively. The bond lengths between carbon atoms, $d_{CC1}$ and $d_{CC2}$, between carbon and hydrogen atoms, $d_{CH}$, the vertical deviation within the octagon, $d_{z1}$, the total vertical deviation, $d_{z2}$, and the bent angle, $\alpha$, are denoted in (A). The energy levels of HOMO and LUMO of the neutral COT molecule with (D) $D_{2d}$, (E) $D_{4h}$, and (F) $D_{8h}$ symmetry are shown, where occupied orbitals are denoted in blue, while unoccupied states are in red. The energy of the Dirac points of pristine graphene is set to zero. In (F), the HOMO and LUMO states are degenerate due to the $D_{8h}$ symmetry.

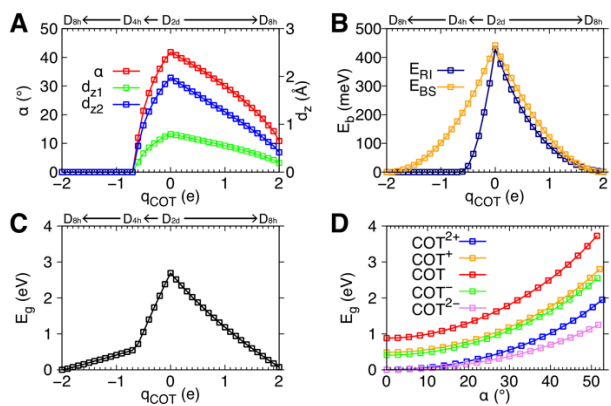

**Fig. 2. Charge dependence of the atomic and electronic properties of a freestanding COT molecule.** (A) The structural parameters, $\alpha$, $d_{z1}$, and $d_{z2}$, are plotted as functions of the charge of the molecule, $q_{COT}$. (B) The energy barriers for the ring inversion $E_{RI}$ and bond shift $E_{BS}$ are plotted as functions of $q_{COT}$ (C) The energy difference between HOMO and LUMO states, $E_g$, are plotted as a function of $q_{COT}$. (D) The energy differences $E_g$ of the dication, the monovalent cation, the neutral molecule, the monovalent anion, and the dianion are plotted as functions of $\alpha$.

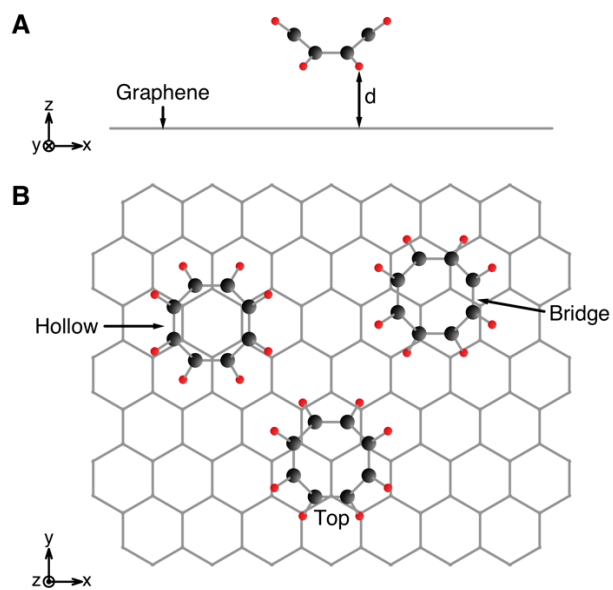

**Fig. 3. Atomic structure of COT molecule adsorbed onto graphene.** (A) Side and (B) Top view. The distance between the molecule and graphene, *d*, is denoted in (A). The adsorbed COT molecules at the hollow, bridge, and top sites are presented in (B). The center of mass of the molecule is vertically located right on the center of a hexagon of graphene in the hollow case, midpoint of a $\pi$-bond of graphene in the bridge case, and on top of a carbon atom of graphene in the top case.

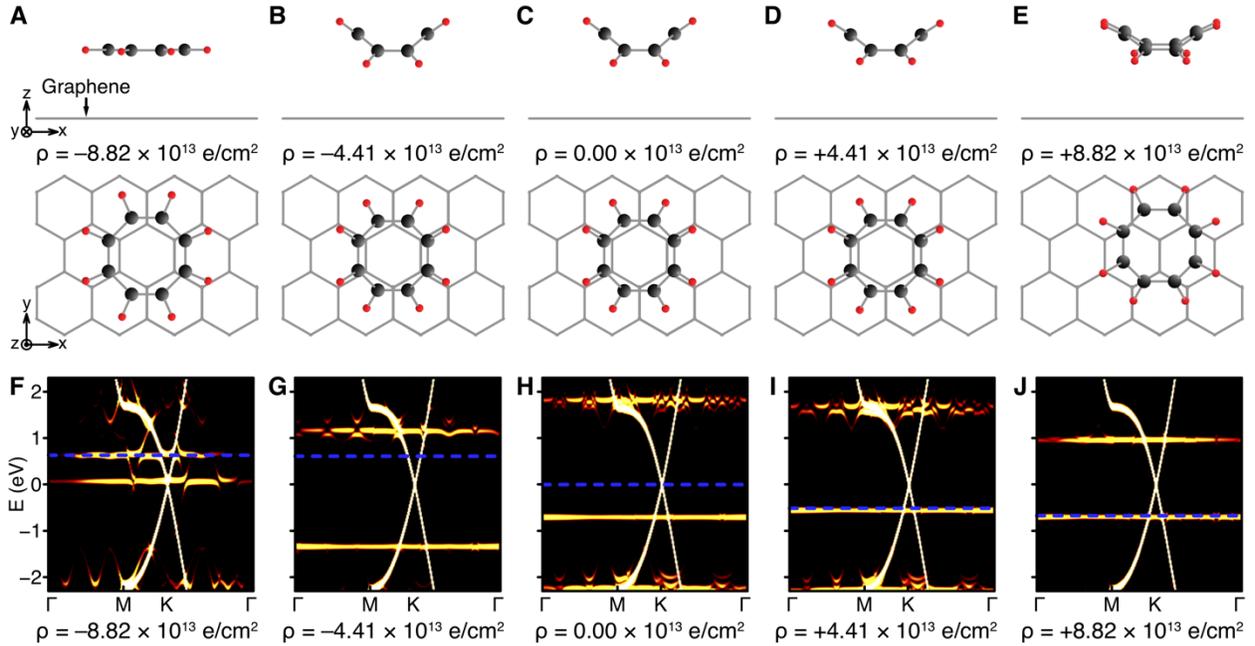

**Fig. 4. The atomic and electronic structures of COT molecule on graphene with various carrier density.** (A-E) Atomic structures with the carrier density of (A) $-8.82\times10^{13}$, (B) $-4.41\times10^{13}$, (C) 0, (D) $+4.41\times10^{13}$, and (E) $+8.82\times10^{13}$ e/cm$^2$ are shown. Top and side views are presented on the top and bottom row, respectively. (F-J) The band structures with the carrier density, corresponding to the atomic structures shown (A-E) in order, are shown. The bands are unfolded with respect to the unit cell of graphene. For each carrier density, the energy of Dirac points is set to zero and the Fermi energy is denoted by blue dashed lines.

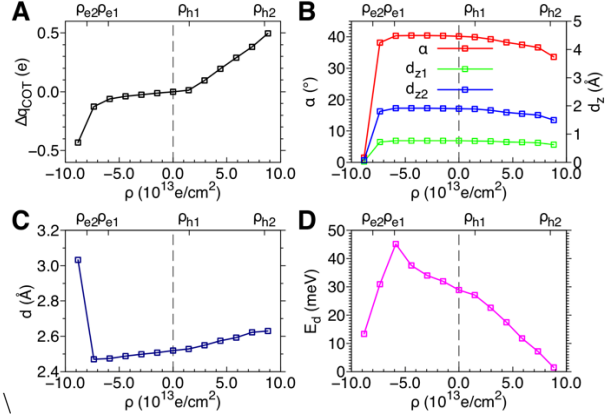

**Fig. 5. Carrier density dependence of a COT molecule adsorbed onto graphene.** (A) The charge transferred to the molecule, $\Delta q_{COT}$, (B) the vertical deviations from planar geometry, $d_{z1}$ and $d_{z2}$, (C) the distance between the molecule and graphene, $d$, and (D) the diffusion barrier, $E_d$, are plotted as functions of the carrier density, $\rho$.

**Tab. 2. The properties of a COT molecule adsorbed onto graphene with various carrier density obtained from DFT calculations.** The charge transferred to the molecule $\Delta q_{COT}$, binding energy $E_b$, and diffusion barrier $E_d$, with various carrier density are presented. The structural properties of the molecule, such as the bond lengths between carbon atoms, $d_{CC1}$ and $d_{CC2}$, between carbon and hydrogen atoms, $d_{CH}$, the vertical deviation within the octagon, $d_{z1}$, the total vertical deviation of the molecule, $d_{z2}$, and the bent angle, $\alpha$, as denoted in Fig. 1(A), are presented as well as the distance between the molecule and graphene, $d$, as denoted in Fig. 2(A).

| $\rho$ (e / cm$^2$) | $\Delta q_{COT}$ (e) | $E_b$ (eV) | $E_d$ (meV) | $d_{CC1}$ (Å) | $d_{CC2}$ (Å) | $d_{CH}$ (Å) | $d_{z1}$ (Å) | $d_{z2}$ (Å) | $\alpha$ (°) | $d$ (Å) |
|---|---|---|---|---|---|---|---|---|---|---|
| +8.82×10$^{13}$ | 0.50 | 1.145 | 1.54 | 1.385 | 1.476 | 1.119 | 0.629 | 1.502 | 33.62 | 2.631 |
| +7.35×10$^{13}$ | 0.38 | 0.983 | 7.15 | 1.382 | 1.479 | 1.119 | 0.691 | 1.673 | 36.67 | 2.623 |
| +5.88×10$^{13}$ | 0.29 | 0.854 | 11.8 | 1.379 | 1.482 | 1.119 | 0.708 | 1.727 | 37.50 | 2.593 |
| +4.41×10$^{13}$ | 0.20 | 0.761 | 17.5 | 1.377 | 1.484 | 1.119 | 0.723 | 1.773 | 38.22 | 2.575 |
| +2.94×10$^{13}$ | 0.10 | 0.708 | 22.6 | 1.375 | 1.488 | 1.120 | 0.745 | 1.844 | 39.26 | 2.550 |
| +1.47×10$^{13}$ | 0.01 | 0.696 | 27.1 | 1.373 | 1.490 | 1.120 | 0.761 | 1.893 | 39.98 | 2.528 |
| 0 | 0.00 | 0.704 | 28.9 | 1.373 | 1.490 | 1.120 | 0.765 | 1.904 | 40.17 | 2.519 |
| −1.47×10$^{13}$ | -0.01 | 0.722 | 32.0 | 1.373 | 1.490 | 1.120 | 0.769 | 1.914 | 40.37 | 2.507 |
| −2.94×10$^{13}$ | -0.02 | 0.751 | 34.0 | 1.373 | 1.490 | 1.120 | 0.770 | 1.917 | 40.42 | 2.498 |
| −4.41×10$^{13}$ | -0.03 | 0.790 | 37.6 | 1.373 | 1.490 | 1.120 | 0.771 | 1.920 | 40.48 | 2.488 |
| −5.88×10$^{13}$ | -0.06 | 0.840 | 45.1 | 1.373 | 1.490 | 1.120 | 0.769 | 1.916 | 40.39 | 2.475 |
| −7.35×10$^{13}$ | -0.13 | 0.912 | 30.9 | 1.376 | 1.488 | 1.120 | 0.725 | 1.806 | 38.19 | 2.470 |
| −8.82×10$^{13}$ | -0.43 | 1.068 | 13.4 | 1.395 | 1.474 | 1.118 | 0.028 | 0.076 | 1.55 | 3.033 |

# Supplement materials for

# Simulating the nanomechanical response of cyclooctateraene molecules on a graphene device

Sehoon Oh[1,2], Michael F. Crommie[1,2,3] and Marvin L. Cohen[1,2]*

[1]*Department of Physics, University of California at Berkeley, Berkeley, CA 94720, USA*

[2]*Materials Sciences Division, Lawrence Berkeley National Laboratory, Berkeley, CA 94720, USA*

[3]*Kavli Energy Nano Sciences Institute at the University of California Berkeley and the Lawrence Berkeley National Laboratory, Berkeley, CA 94720, USA*

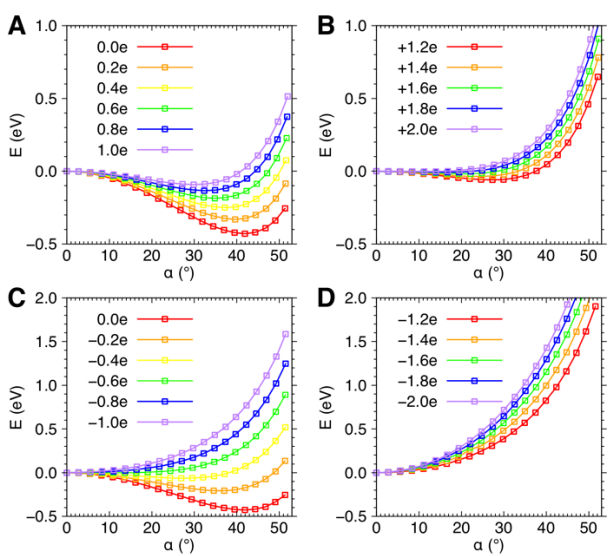

**Fig. S1. Energy as functions of the bent angle and charge of a freestanding COT molecule.** The total energies are plotted as functions of the bent angle (A,B) for positively and (C,D) negatively charged molecule.

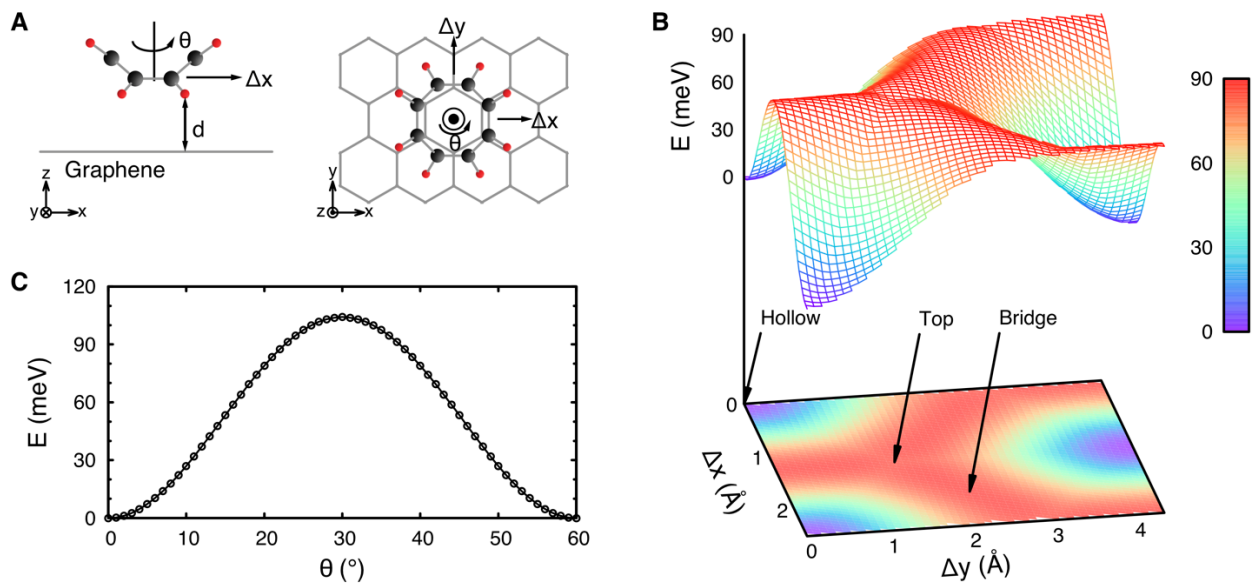

**Fig. S2. Energy as functions of the parallel shift and rotation of a COT molecule adsorbed onto graphene.** (A) The atomic structure of a COT molecule with a tub-shaped conformation adsorbed onto graphene is schematically illustrated. Side and top views are presented. The atomic structure is constructed using the atomic positions of the freestanding COT molecule and pristine graphene with a distance between the molecule and graphene $d = 2.519$ Å. The total energies are obtained as functions of (B) parallel shift, $\Delta x$ and $\Delta y$, and (C) the rotational angle, $\theta$, without further relaxing, where the origin, $\Delta x = \Delta y = \theta = 0$, is set to the position and orientation of the molecule shown in (A), and $E(\Delta x = \Delta y = \theta = 0)$ is set to zero.

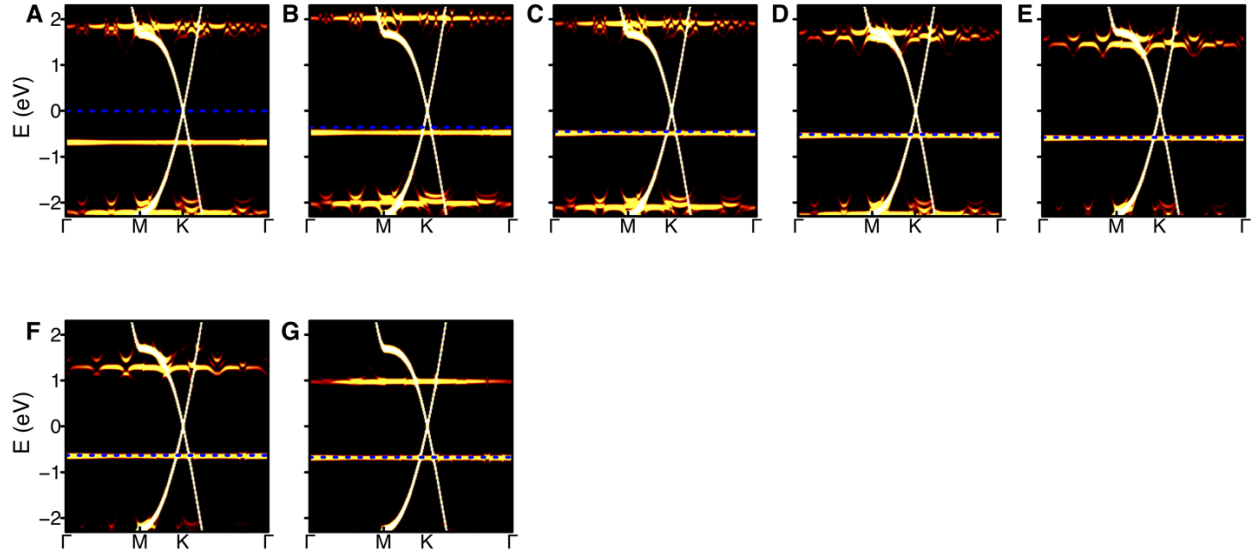

**Fig. S3. The electronic band structures of a COT molecule on graphene with hole carrier density.** The band structures with the carrier density of (A) 0, (B) +1.47×10$^{13}$, (C) +2.94×10$^{13}$, (D) +4.41×10$^{13}$, (E) +5.89×10$^{13}$, (F) +7.35×10$^{13}$, and (G) +8.82×10$^{13}$ e/cm$^2$ are shown, from left to right. The bands are unfolded with respect to the unit cell of graphene. The energies of Dirac points are set to zero and the Fermi energies are denoted by blue dashed lines.

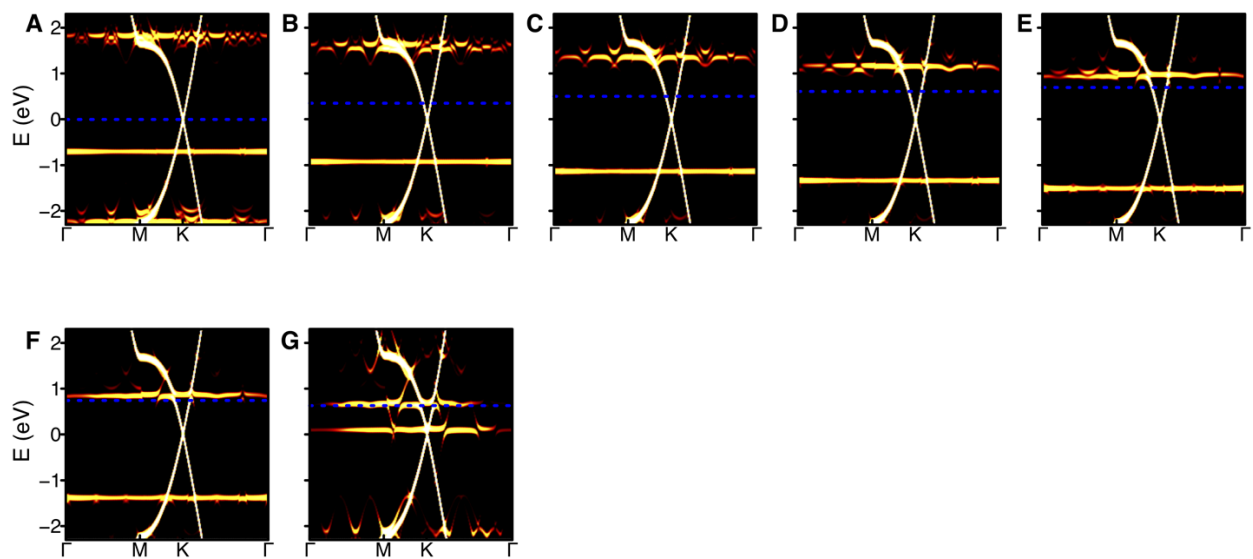

**Fig. S4. The electronic band structures of a COT molecule on graphene with electron carrier density.** The band structures with the carrier density of (A) 0, (B) $-1.47\times10^{13}$, (C) $-2.94\times10^{13}$, (D) $-4.41\times10^{13}$, (E) $-5.89\times10^{13}$, (F) $-7.35\times10^{13}$, and (G) $-8.82\times10^{13}$ e/cm$^2$ are shown. The bands are unfolded with respect to the unit cell of graphene. The energies of Dirac points are set to zero and the Fermi energies are denoted by blue dashed lines.